\documentclass[11pt,preprint]{aastex}
\usepackage{graphicx}

\begin{document}
%\begin{article}
%\begin{opening}
\title{Eclipsing Binaries Showing Light-Time Effect}
%\author{Petr \surname{Zasche}\thanks{E-mail address: Petr.Zasche@Email.cz}}
%\runningauthor{Petr Zasche}
%\runningtitle{Eclipsing Binaries Showing Light-Time Effect}

\author{Petr Zasche}
\affil{Astronomical Institute, Faculty of Mathematics and Physics, \\
Charles University, CZ-180 00 Prgue 8, V Hole\v{s}ovi\v{c}k\'ach
2, Czech Republic} \email{zasche@sirrah.troja.mff.cuni.cz}

%\institute{Astronomical Institute, Charles University, Prague,Czech Republic}
%\date{April 15, 2004}

\begin{abstract} Four eclipsing binaries which show
apparent changes of period have been studied in case of possible
presence of the light-time effect. With the least squares method
we calculated new light elements of these systems and mass
function of predicted third body and its minimum mass. We
discussed probability of a presence of such bodies in case of mass
function, changes in radial velocity and third light in solution
of light curves, respectively.
\end{abstract}

%\end{opening}

\section{Introduction}
Light-time effect (hereafter LITE) in eclipsing binaries, caused
by the orbital motion of the eclipsing pair around the barycenter
of the triple system, produces period variation in minima timings.
It was explained by Irwin in 1959 and its necessary criteria have
been mentioned by Friebos-Conde \& Hertzeg in 1973. They are
following: agreement with a theoretical LITE curve, secondary
minima behave identically with primary ones, reasonable value of
the mass function and corresponding variations in radial velocity,
respectively.

We found, that most of the necessary criteria listed above are
satisfied in all systems presented below. Their $O\!-\!C$ diagrams
are also presented. The curves represent computed LITE versus
epoch. The individual primary and secondary minima are denoted by
dots and circles, respectively. Larger symbols correspond to the
photoelectric or CCD measurements which were given higher weights
in our computations. All data were found in published papers.

\begin{figure}[t!]
% \hspace{12mm}
\begin{center}
 \scalebox{0.68}{\includegraphics{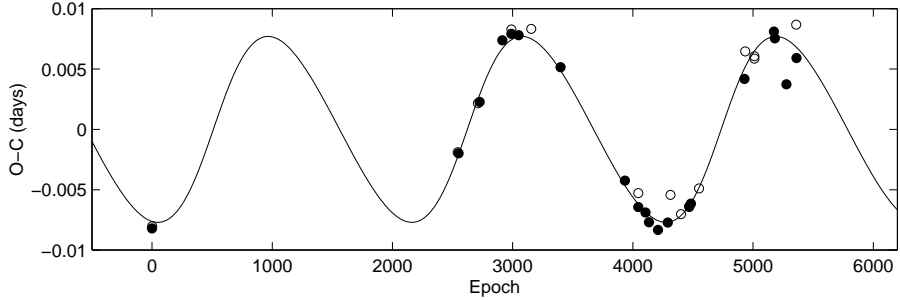}}
 \caption[]{The O-C diagram of AR Aur.}
 \label{ARAur}
 \end{center}
\end{figure}

\section{Observations}
 \underline{\bfseries AR Aur} (HD 34364) as a variable was
discovered in 1931. For the computations here we use only accurate
photoelectric and CCD minima timings. From these data we have
calculated new light elements
\[ \mathrm{Min\:I} = \mathrm{HJD}\;\:24\;27887,7299 +
 4,\!\!^{\rm d}13466577 \cdot {\rm E}.\]
The period investigation in 1988 was made by Chochol et al. With
new minima timings we were able to determine third body orbit with
more plausibility. Because of the relative low mass of the
unresolved component, the third light must be less than 1 \% of
the total light. Also the variations in radial velocities, because
of the low mass and errors in velocity measurements, are nowadays
still inconclusive.

\begin{figure}[b!]
 \begin{center}
 \scalebox{0.68}{\includegraphics{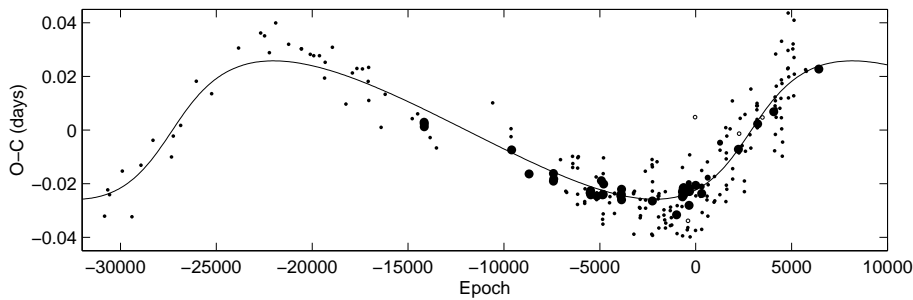}}
 \caption[]{O-C diagram of R CMa}
 \label{RCMa}
 \end{center}
\end{figure}

%\smallskip
 \underline{\bfseries R CMa} is one of the most exemplary
eclipsing binary system which show LITE. Almost 300 minima times
of R CMa (HD 57167) were collected. Our computations with these
data lead to new light elements
\[ \mathrm{Min\:I}
= \mathrm{HJD}\;\:24\;45391,2570 + 1,\!\!^{\rm d}13594210 \cdot
 {\rm E}.\]
We can discuss our results with system analysis from Ribas et al.
(2002), who calculated lower $p_3$ but higher semiamplitude, which
gives higher mass function. Variations in radial velocities and
also astrometric confirmation from Hipparcos satellite is still
inconclusive. The third body is probably WD or M-type dwarf.

\begin{figure}[t]
 \begin{center}
 \scalebox{0.68}{\includegraphics{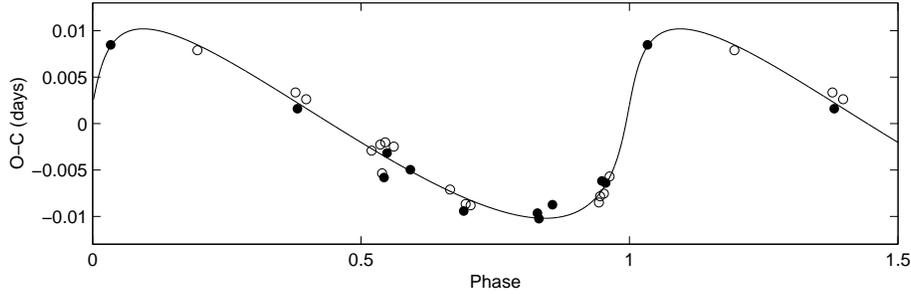}}
 \caption[]{The O-C phase diagram of FZ CMa.}
 \label{FZCMa}
 \end{center}
\end{figure}

%\smallskip
 \underline{\bfseries FZ CMa}. Moffat and Vogt performed detailed
spectroscopic and photometric study of FZ CMa (HD 52942) in 1983,
but period investigation was realized only with minima from 1973
to 1980. New light elements are
\[ \mathrm{Min\:I} =
\mathrm{HJD}\;\:24\;41743,5915 + 1,\!\!^{\rm d}27303611 \cdot
 {\rm E}.\]
As we can see in Table I, minimum mass of this body is very high
and in very eccentric orbit, so in the system would dominate, what
seems unlikely.  Possible solution is, that the third body is a
binary too, what seems likely and nearly corresponds to the
spectral analysis.

\begin{figure}[b]
 \begin{center}
 \scalebox{0.68}{\includegraphics{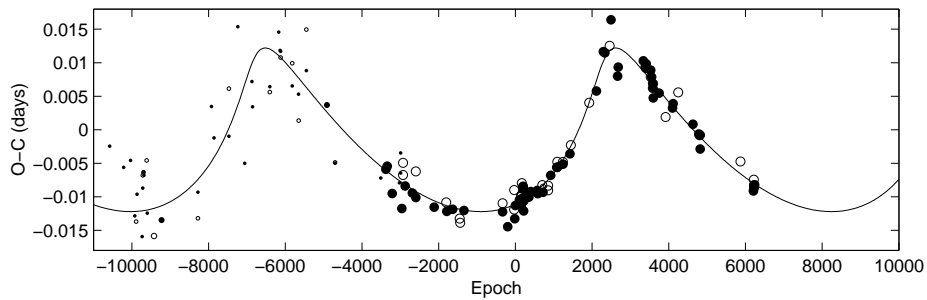}}
 \caption[]{The O-C diagram of TX Her}
 \label{TXHer}
 \end{center}
\end{figure}

%\smallskip
 \underline{\bfseries TX Her} (HD 156965) as a variable
was found by Miss Cannon (see Pickering 1910). It was very often
observed, what leads to large list of visual  minima times and
from 40's to accurate photoelectric and CCD ones, which leads to
new light elements
\[ \mathrm{Min\:I} = \mathrm{HJD}\;\:24\;40008,3686 +
 2,\!\!^{\rm d}05980975 \cdot {\rm E}.\]
We can compare our results with the latest LITE study of TX Her by
Ak et al. (2004), who calculated higher value of excentricity, and
semiamplitude, and lower period $p_3$, so their mass function is
nearly double than our.

 In table below are $M_i$ masses of components, $p_3$ period of the
third body, $A$ semiamplitude of LITE (see e.g. Mayer 1990), $e$
excentricity, $\omega$ length of periastron, $f(M_3)$ computed
mass function and $M_{3,min}$ minimum mass (for $i$ = 90$^\circ$)
of this predicted body, respectively.

\begin{table}[h]
 \begin{center}
 \label{tabl}
 \caption{The results for four studied LITE systems.}
 \scalebox{0.75}{
 \begin{tabular}{cccccccccc}
 \hline \hline
Name of star&      Spectrum   &   $ M_1$     &  $ M_2$      &  $ p_3 $       &    $ A $     &    $e$     & $  \omega $ & $  f(M_3)  $    & $ M_{3,min}   $ \\
            &                 &   [$M_\odot$] &   [$M_\odot$] &   [years]      &    [days]    &            &    [deg]    &   [$M_\odot$]    &   [$M_\odot$]   \\
\hline
AR Aur   &         B9V+B9.5V    &    2.48     &     2.29    &     23.92      &      0.0077      &   0.209 &      2.3     &   0.0044         &     0.497       \\
FZ CMa   &   B2.5IV-V+B2.5IV-V  &    5.40     &     5.30    &     1.511      &      0.0102      &   0.794 &      7.0     &   10.338         &     22.51      \\
 R CMa   &         F0V+K2IV     &    1.07     &     0.17    &     93.89      &      0.0258      &   0.498 &     357.8    &   0.0155         &     0.338      \\
TX Her   &          A7+F0       &    1.62     &     1.45    &     51.53      &      0.0122      &   0.654 &     48.6     &   0.0048         &     0.387      \\
\hline \hline
\end{tabular}}
 \end{center}
\end{table}
\vspace{-8mm}
\section{Conclusion}
We have derived new LITE parameters for four eclipsing binaries by
their $O\!-\!C$ diagram analysis. But for the final decision of
the presence of LITE we need further accurate timings of minima or
appropriate confirmation by another method (radial velocity
variations, third light in a light curve analysis or astrometric
observations).

\section{Acknowledgements}
This research has made use of the SIMBAD database, operated at
CDS, Strasbourg, France, and of NASA's Astrophysics Data System
Bibliographic Services. We are grateful to P. Niarchos et.al. for
their minima measurement of FZ CMa, which wasn't still published.
This investigation was supported by the Grant Agency of the Czech
Republic, grant No. 205/04/2063.

%\end{article}
\end{document}